\documentclass[a4paper]{jpconf}
\usepackage{graphicx}
\begin{document}
\title{From SICs and MUBs to Eddington}

\author{Ingemar Bengtsson}

\address{Fysikum, Stockholms Universitet, S-106 91 Stockholm, Sweden}

\ead{ingemar@physto.se}

\begin{abstract}
This is a survey of some very old knowledge about Mutually Unbiased Bases 
(MUB) and Symmetric Informationally Complete POVMs (SIC). In prime 
dimensions the former are closely tied to an elliptic normal curve 
symmetric under the Heisenberg group, while the latter are believed to 
be orbits under the Heisenberg group in all dimensions.  In dimensions 
3 and 4 the SICs are understandable in 
terms of elliptic curves, but a general statement escapes 
us. The geometry of the SICs in 3 and 4 dimensions is discussed in some detail. 
\end{abstract}

\section{Introduction}

Two problems that have attracted much attention in the quantum 
information community are the MUB and SIC problems for Hilbert spaces of 
finite dimension $N$. In the MUB problem \cite{Ivanovic, Wootters1} 
one looks for $N+1$ orthonormal bases that are mutually unbiased, in the sense 
that 

\begin{equation} |\langle e_m|f_n\rangle|^2 = \frac{1}{N} \ , \hspace{8mm} 
0 \leq m,n \leq N-1 \ , \end{equation}

\noindent whenever the vector $|e_m\rangle$ belongs to one basis and the 
vector $|f_n\rangle$ to another. In the SIC problem \cite{Zauner, Renes} 
one looks for a symmetric and informationally complete POVM, which translates 
to the problem of finding $N^2$ unit vectors $|\psi_i\rangle$ such that 

\begin{equation} |\langle \psi_i|\psi_j\rangle |^2 = \frac{1}{N+1}
\ , \hspace{8mm} 0 \leq i,j \leq N^2 - 1 \ , \end{equation}

\noindent whenever $i \neq j$. These problems are hard. For the MUB problem 
an elegant solution exists whenever $N$ is a power of a prime \cite{Wootters2}. 
For the SIC problem quite ad hoc looking analytic solutions are known for 
eighteen different dimensions; these are described (and in some cases derived) 
by Scott and Grassl, who also give full references to the earlier literature 
\cite{Grassl}. The belief in the community is that a complete set of $N+1$ 
MUB does not exist for general $N$, while the SICs do. 

Since the problems are so easy to state, it is not surprising that they have 
been posed independently in many different branches of science. One purpose 
of this article is to describe what nineteenth century geometers had to say 
about them. A story told by Eddington \cite{Eddington} 
is relevant here: 

\

\noindent {\small Some years ago I worked out the structure of this group of operators 
in connection with Dirac's theory of the electron. I afterwards learned that 
a great deal of what I had written was to be found in a treatise on Kummer's 
quartic surface. There happens to be a model of Kummer's quartic surface in 
my lecture-room, at which I had sometimes glanced with curiosity, wondering 
what it was all about. The last thing that entered my head was that I had 
written (somewhat belatedly) a paper on its structure. Perhaps the author 
of the treatise would have been equally surprised to learn that he was 
dealing with the behaviour of an electron.}

\

\noindent We will see what Eddington saw as we proceed. Meanwhile, let us 
observe that when $N$ is a prime 
the MUB are the eigenbases of the $N+1$ cyclic subgroups of the Heisenberg 
group, while there is a conjecture (enjoying very considerable numerical 
support \cite{Grassl}) that the SICs can always be chosen to be special orbits of 
this group. When $N$ is a power of a prime the solution of the MUB problem 
shifts a little, since the MUBs now consist of eigenvectors of the cyclic subgroups 
of the Heisenberg group defined over a finite field rather than over the 
ring of integers modulo $N$. Concerning SICs that are orbits under the 
Heisenberg group there is a link to the MUB problem: If the dimension 
$N$ is a prime the SIC Bloch vectors, when projected onto any one of the 
MUB eigenvalue simplices, have the same length for all the 
$N+1$ MUB \cite{Khat, ADF}. 

In mathematics elliptic curves provide the natural home for the Heisenberg 
group, so it seems natural to investigate if elliptic curves can be used 
to illuminate the MUB and SIC problems. In dimensions 3 \cite{Hughston} and 
4 they certainly can, as we will see, but in higher dimensions I am not so 
sure. There will be some comments and formulas that I could not find in 
the books and papers I studied, but keeping 
Eddington's example in mind I do not claim originality for them.
 
\section{Two pieces of background information}
We had better define the Heisenberg group properly. A defining non-unitary 
representation is given by the upper triangular matrices 

\begin{equation} g(\gamma, \alpha, \beta) = 
\left( \begin{array}{ccc} 1 & \alpha &  \gamma \\ 0 & 1 & \beta \\ 
0 & 0 & 1 \end{array} \right) \ . \end{equation}

\noindent Here the matrix elements belong to some ring. In the original 
Weyl-Heisenberg group \cite{Weyl} they are real numbers, but here we are 
more interested in the case that they belong to the ring of integers 
modulo $N$. We denote the resulting group by $H(N)$. It is generated 
by two elements $X$ and $Z$ obeying 

\begin{equation} ZX = qXZ \ , \hspace{8mm} X^N = Z^N = {\bf 1} \ , 
\hspace{8mm} q = e^{\frac{2\pi i}{N}} \ . \end{equation}

\noindent For $N = 2$ we can use the Pauli 
matrices to set $X = \sigma_X$, $Z = \sigma_Z$, which makes it possible 
to remember the notation. We will consider the group projectively, so for 
our purposes it can often be regarded as a group of order $N^2$. 

Because $q$ is a primitive $N$th root of unity the unitary representation 
in which $Z$ is diagonal is unique up to permutations \cite{Weyl}. 
It is known as the clock and shift representation. If the components of 
any vector are denoted $x_a$ the action is given by

\begin{equation} \begin{array}{lll} X: & & x_0 \rightarrow x_{N-1} \rightarrow 
x_{N-2} \rightarrow \dots \rightarrow x_1 \rightarrow x_0 \\
\ \label{group} \\ Z: & & x_a \rightarrow q^ax_a \end{array} \ , 
\hspace{8mm} 0 \leq a \leq N-1 \ . \end{equation}

\noindent The unitary automorphism group of the Heisenberg group plays 
prominent roles in quantum information theory \cite{Fivel, Gottesman}, 
and is often called the Clifford group. In the older literature the 
Heisenberg group is sometimes called the Clifford collineation group, 
and the Clifford group is called the Clifford transform group \cite{Horadam}. 
Although we will discuss it in detail for the case $N = 3$ later on, we will 
mostly be concerned with automorphisms of order 2. In the clock and shift 
representation such an automorphism acts according to 

\begin{equation} A: \ \ \  x_a \leftrightarrow x_{-a} \ . \label{A} \ . 
\end{equation}

\noindent Adding this generator leads us to consider an extended group which is 
twice as large as $H(N)$. In quantum information language the involution $A$ is 
generated by one of Wootters' phase point operators \cite{Wootters1}. 
Finally there is the curious conjecture \cite{Zauner} that 
the SIC vectors are always left invariant by a unitary automorphism of the 
Heisenberg group having order 3. No one knows why this should be so, 
but it does appear to be true \cite{Marcus, Grassl}, and in four dimensions 
we will see exactly how it happens.  

What is special about the case when $N$ is prime is that $H(N)$ then admits 
$N+1$ cyclic subgroups of order $N$, forming a flower with $N+1$ petals 
with only the unit element in common. Correspondingly there are $N+1$ 
eigenbases, and they necessarily form a complete set of 
MUB \cite{Vatan}. In prime power dimensions $N = p^k$ the known complete set 
of MUB is the 
set of eigenbases of the cyclic subgroups of a Heisenberg group defined 
over a Galois field. The only case we will discuss is when $N = 4$, for 
which the Galois Heisenberg group is the tensor product $H(2) \otimes H(2)$. 

Another piece of background information is that SICs and MUBs look 
natural in Bloch space, which is the $N^2-1$ dimensional 
space of Hermitean operators of trace 1, considered as a vector space with 
the trace inner product and with the maximally mixed state at the origin. 
Density matrices form a convex body in Bloch space. A SIC is simply a regular 
simplex in Bloch space, inscribed into this convex body. But it is not easy 
to rotate the simplex while keeping the body of density matrices fixed, 
because the symmetry group of this body is only $SU(N)/Z_N$, a rather 
small subgroup of $SO(N^2-1)$ as soon as $N > 2$. This is why the SIC 
problem is hard. An orthonormal basis is a regular simplex with only 
$N$ corners, spanning some $(N-1)$-plane through the origin in Bloch space. 
Two bases are mutually unbiased if the corresponding $(N-1)$-planes are 
totally orthogonal, from which it immediately follows that no more than 
$N+1$ MUB can exist.

Any pure state corresponds to a Bloch vector of a definite length. Given a 
complete set of MUB we can project this vector onto the $N+1$ different 
$(N-1)$-planes defined by the MUB. Should it happen that these projected 
vectors all have the same length the vector is as it were unbiased with 
respect to the MUB, and is then---for some reason---called a 
Minimum Uncertainty State \cite{Wootters3, Appleby}. The condition on a 
state vector to be unbiased in this sense is easily worked out using the 
Euclidean metric on Bloch space in conjunction with Pythagoras' theorem. 
Choose any one of the MUB as the computational basis, and express the 
Hilbert space components of a unit vector with respect to that basis as  

\begin{equation} x_a = \sqrt{p_a}e^{i\mu_a} \ , \hspace{8mm} 
\sum_{n = 0}^{N-1} p_a = 1 \ . \label{octant} \end{equation}

\noindent If the corresponding Bloch vector projected onto the $(N-1)$-plane spanned by 
the computational basis has the length appropriate to a Minimum Uncertainty 
State it must be true that 

\begin{equation} \sum_{a = 0}^{N-1}p_a^2 = \frac{2}{N+1} \ . \label{MUS} 
\end{equation} 

\noindent This is simple enough, but there is the complication that this has 
to be done for all the $N+1$ MUB, which will give an additional set of $N$ 
constraints on the phases ensuring that the vector has the appropriate 
length when projected to the other MUB planes. We spare the reader from 
the details, but we repeat that all Heisenberg covariant SIC vectors are 
Minimum Uncertainty States whenever $N$ is a prime. Examining the 
proof of this interesting statement shows that something similar 
is true also when no complete set of MUB is available: In any eigenbasis 
of a cyclic subgroup of $H(N)$ of order $N$ eq. (\ref{MUS}) will hold 
for any vector belonging to a Heisenberg covariant SIC \cite{Khat, ADF}. 
This is true regardless of how many bases of this kind there are.   

\section{The syzygetic Hesse pencil}
We now descend to the complex projective plane, and begin by introducing the 
language used by nineteenth century geometers. Points are represented by ket 
or column vectors in ${\bf C}^3$, or more precisely by one-dimensional 
subspaces, while lines are represented by two-dimensional subspaces. Using 
the scalar product in Hilbert space we can equally well represent the 
lines by bra or row vectors orthogonal to the subspaces they represent, 
so that the relation  

\begin{equation} \langle Y|X\rangle = 0 \label{ett} \end{equation}

\noindent means that the point $X$ lies on the line $Y$. The two-dimensional 
subspace representing the line consists of all vectors whose scalar product with 
the bra vector $\langle Y|$ vanishes. Since there is a one-to-one correspondence 
$|X\rangle \leftrightarrow \langle X|$ between bras and kets there is also a 
one-to-one correspondence between points and lines. Clearly eq. 
(\ref{ett}) implies that 

\begin{equation} \langle X|Y\rangle = 0 \ , \end{equation}

\noindent which says that the point $Y$ lies on the line $X$. This is known as 
the duality between points and lines in the projective plane. 

We will study complex plane curves defined by homogeneous polynomials 
in three variables. Linear polynomials define two-dimensional 
subspaces, that is to say two real-dimensional subsets of the complex plane, 
and by the above they define projective lines. Intrinsically they are 
spheres, namely Bloch spheres, because ${\bf CP}^1 = {\bf S}^2$. 
Quadratic polynomials or quadrics define conic sections, and over the 
complex numbers the intrinsic geometry of a conic section is again that of a 
sphere. The set of spin coherent states is an example \cite{BH}. To 
the next order in complication we choose a cubic polynomial. We require the curve 
to transform into itself under the Heisenberg group in the clock and shift 
representation (\ref{group}). Up to an irrelevant overall constant the most 
general solution for the cubic is then  

\begin{equation} P = x^3 + y^3 + z^3 + txyz \ . \label{cubic} \end{equation}

\noindent Here $t$ is a complex number parametrising what is known as the 
syzygetic Hesse pencil of cubics. Intrinsically each cubic is a torus rather 
than a sphere. We observe that the polynomial is automatically invariant 
under the additional involution $A$ given above in (\ref{A}). 

Hesse \cite{Hesse}, and before him Pl\"ucker \cite{Plucker}, studied this family 
of curves in detail. Their first object was to determine the inflection points. 
They are given by those points on the curve for which the determinant of its 
matrix of second derivatives---its Hessian---vanishes. In the present case this 
is a cubic polynomial as well; in fact  

\begin{equation} H = \det{\partial_i\partial_jP} = 
(6^3 + 2t^3)xyz - 6t^2(x^3 + y^3 + z^3) \ . \end{equation}

\noindent This is again a member of the Hesse pencil of cubics. In astronomy 
a ``syzygy'' occurs when three planets lie on a line, so we can 
begin to appreciate why the pencil is called ``syzygetic''. The inflection 
points are given by $P = H = 0$. By B\'ezout's theorem two cubics in the 
complex projective plane intersect in nine points, hence there are nine 
inflection points. They coincide for all cubics in the pencil, and are 
given by 

\begin{equation} \left[ \begin{array}{ccccccccc} 0 & 0 & 0 & -1 & - q & - q^2 & 
1 & 1 & 1 \\ 1 & 1 & 1 & 0 & 0 & 0 & -1 & - q & - q^2 \\ -1 & - q & - q^2 & 
1 & 1 & 1 & 0 & 0 & 0 \end{array} \right] \ . \label{points} \end{equation}

\noindent This is recognisable as a set of nine 
SIC vectors covariant under the Heisenberg group \cite{Zauner, Renes}. We can 
normalise our vectors if we want to, but in the spirit of projective geometry 
we choose not to.

There are four singular members of the Hesse pencil, defined by values of the 
parameter $t$ such that there are non-zero solutions to $P = P_{,x} = P_{,y} 
= P_{,z} = 0$. These values are 

\begin{equation} t = \infty \hspace{5mm} \mbox{and} \hspace{5mm} t^3 = - 
3^3 \ . \end{equation}

\noindent If $t = \infty$ the polynomial reduces to $xyz = 0$. In 
this case the singular cubic consists of three projective lines that make 
up a triangle. The remaining three singular cases will give rise to three 
other triangles. Therefore the syzygetic pencil singles 
out 4 special triangles in the projective plane, given by their 12 vertices 

\begin{eqnarray} \triangle^{(0)} = \left[ \begin{array}{ccc} 1 & 0 & 0 \\ 
0 & 1 & 0 \\ 0 & 0 & 1 \end{array} \right] \ , \hspace{6mm} 
\triangle^{(1)} = \left[ \begin{array}{ccc} 1 & q^2 & q^2 \\ q^2 & 1 & q^2 \\ 
q^2 & q^2 & 1 \end{array} \right] \ ,  \hspace{12mm} \nonumber \\ 
\label{MUB3} \\
\triangle^{(2)} = \left[ \begin{array}{ccc} 1 & q & q \\ q & 1 & q \\ 
q & q & 1 \end{array} \right] \ , \hspace{8mm} 
\triangle^{(\infty )} = \left[ \begin{array}{ccc} 1 & 1 & 1 \\ 1 & q & q^2 \\ 
1 & q^2 & q \end{array} \right] \ , \hspace{12mm} \nonumber \end{eqnarray}

\noindent where $q = e^{2\pi i/3}$. The columns, labelled consecutively by 
$0,1,2$, can indeed be regarded as 12 points or by duality as 12 lines. 
The four triangles are referred to as the inflection triangles.

What gives the triangles their name is the remarkable fact that the nine 
inflection points lie by threes on their twelve edges. Hesse calls this 
a ``{\it sch\"onen Lehrsatz}'', and attributes it to Pl\"ucker \cite{Plucker}. 
It is not hard 
to verify. After a small calculation one finds that the orthogonalities 
between the columns in the four triangles and the vectors representing 
the inflection points are as follows:
 
\

{\tiny \begin{tabular}{|c||ccc|ccc|ccc|ccc|} \hline 
\  & $\triangle_0^{(0)}$ & $\triangle_1^{(0)}$& $\triangle_2^{(0)}$ & $\triangle_0^{(1)}$ 
& $\triangle_1^{(1)}$ & $\triangle_2^{(1)}$ & $\triangle_0^{(2)}$ & $\triangle_1^{(2)}$ & 
$\triangle_2^{(2)}$ & $\triangle_0^{(\infty )}$ & $\triangle_1^{(\infty )}$ & 
$\triangle_2^{(\infty )}$ \\ \hline \hline 
$X_0$ & $\bullet$ & \ & \ & $\bullet$ & \ & \ & $\bullet$ & \ & \ & $\bullet$ & \ & \ \\
\ & \ & \ & \ & \ & \ & \ & \ & \ & \ & \ & \ & \ \\
$X_1$ & $\bullet$ & \ & \ & \ & \ & $\bullet$ & \ & $\bullet$ & \ & \ & $\bullet$ & \ \\ 
\ & \ & \ & \ & \ & \ & \ & \ & \ & \ & \ & \ & \ \\
$X_2$ & $\bullet$ & \ & \ & \ & $\bullet$ & \ & \ & \ & $\bullet$ & \ & \ & $\bullet$ \\
\ & \ & \ & \ & \ & \ & \ & \ & \ & \ & \ & \ & \ \\ 
$Y_0$ & \ & $\bullet$ & \ & \ & $\bullet$ & \ & \ & $\bullet$ & \ & $\bullet$ & \ & \ \\
\ & \ & \ & \ & \ & \ & \ & \ & \ & \ & \ & \ & \ \\
$Y_1$ & \ & $\bullet$ & \ & $\bullet$ & \ & \ & \ & \ & $\bullet$ & \ & $\bullet$ & \ \\ 
\ & \ & \ & \ & \ & \ & \ & \ & \ & \ & \ & \ & \ \\
$Y_2$ & \ & $\bullet$ & \ & \ & \ & $\bullet$ & $\bullet$ & \ & \ & \ & \ & $\bullet$ \\
\ & \ & \ & \ & \ & \ & \ & \ & \ & \ & \ & \ & \ \\ 
$Z_0$ & \ & \ & $\bullet$ & \ & \ & $\bullet$ & \ & \ & $\bullet$ & $\bullet$ & \ & \ \\
\ & \ & \ & \ & \ & \ & \ & \ & \ & \ & \ & \ & \ \\
$Z_1$ & \ & \ & $\bullet$ & \ & $\bullet$ & \ & $\bullet$ & \ & \ & \ & $\bullet$ & \ \\ 
\ & \ & \ & \ & \ & \ & \ & \ & \ & \ & \ & \ & \ \\
$Z_2$ & \ & \ & $\bullet$ & $\bullet$ & \ & \ & \ & $\bullet$ & \ & \ & \ & $\bullet$ \\
\hline
\end{tabular}}

\

\

\noindent Thus we have 

\begin{equation} \langle \Delta_0^{(0)} |X_0\rangle = 
\langle \Delta_0^{(0)} |X_1\rangle = \langle \Delta_0^{(0)} |X_2\rangle = 
0 \end{equation} 

\noindent and so on. Recalling the interpretation of the vanishing scalar 
products we see by 
inspection of the table that Hesse's beautiful theorem is true.

We have verified that there exists a configuration of 9 points and 12 lines 
such that each point belongs to four lines, and each line goes through three points. 
This is denoted $(9_4, 12_3)$, and is known as the Hesse configuration. Using the 
duality between points and lines we have also proved the existence of the 
configuration $(12_3, 9_4)$. From an abstract point of view such a configuration 
is a combinatorial object known as a finite affine plane \cite{Dolgachev}. In the 
language of quantum information theory the inflection triangles form a complete 
set of four MUB, while the inflection points form a SIC. 

We can now expand on our discussion of group theory in section 2. 
First, every plane cubic can be 
regarded as a commutative group in a natural way. This is not surprising, 
given that the curve is intrinsically a torus---that is a group manifold. 
The idea relies on B\'ezout's theorem, which this time assures us that any 
line intersects the cubic in three points---two of which coincide 
if the line is a tangent, and all of which coincide if the line is a line of 
inflection. An arbitrary point on the cubic is taken to be the identity element, 
and denoted $O$. To add two arbitrary points $A$ and $B$ on the 
cubic, draw the line between them and locate its third intersection point $P$ 
with the cubic. Then draw the line between $O$ and $P$ and again locate the 
third intersection point $C$. By definition then $A + B = C$. All the group 
axioms are obeyed, although it is non-trivial to prove associativity. 

Now choose the origin to sit at one of the inflection points. With Hesse's 
construction in hand one sees that the nine inflection points form a 
group of order nine, which is precisely the projective Heisenberg group.   
This is also the torsion group of the curve, meaning that it contains all 
group elements of finite order. Because they are group elements of order 
3 the inflection points are also called 3-torsion points. 

Next we ask for the group of transformations transforming the cubics of the 
Hesse pencil among themselves. Recall that the parameter $t$ in the Hesse 
cubic (\ref{cubic}) can serve as a complex coordinate on a sphere. The 
four singular members of the pencil defines a tetrahedron 
on that sphere. Transformations within the pencil act as M\"obius transformations 
on the complex number $t$. Moreover they must permute the singular members 
of the Hesse pencil among themselves. This means that they form a well known 
subgroup of $SO(3)$, namely the symmetry group $A_4$ of the regular 
tetrahedron. It enjoys the isomorphism 

\begin{equation} A_4 \sim PSL(2, {\bf F}_3) \ , \end{equation}

\noindent where ${\bf F}_3$ is the field of integers modulo 3. The group 
$SL(2, {\bf F}_3)$ consists of unimodular two by two matrices with 
integer entries taken modulo three; here only its projective part enters 
because the subgroup generated by the matrix $-{\bf 1}$ gives rise to 
the involution $A$ and does not act on $t$, although it does 
permute the inflection points among themselves. The full symmetry 
group of the pencil is a semi-direct product of the Heisenberg group 
and $SL(2, {\bf F}_3)$. This is the affine group on a finite affine 
plane. It is known as the Hessian group \cite{Jordan}, or as the 
Clifford group. 

There are many accounts of this material in the literature, from geometric  
\cite{Grove}, undergraduate \cite{Gibson}, and modern \cite{Artebani} points 
of view. It forms a recurrent theme in Klein's history of nineteenth 
century mathematics \cite{Klein}. The fact that the inflection points form 
a SIC was first noted by Lane Hughston \cite{Hughston}.   

\section{The elliptic normal curve in prime dimensions} 
Felix Klein and the people around him put considerable effort into the 
description of elliptic curves embedded into projective spaces of dimension 
higher than 2. They proceeded by means of explicit parametrisations of 
the curve using Weierstrass' $\sigma$-function \cite{Bianchi, Hulek}. As far 
as we are concerned now, we only need to know that the symmetries they 
built into their curves is again the Heisenberg group supplemented with 
the involution $x_a \leftrightarrow x_{-a}$ coming from the Clifford group. 
An analysis of this group of symmetries leads directly to ``{\it une 
configuration tr\`es-remarquable}'' originally discovered by Segre 
\cite{Segre}. We will present it using some notational improvements 
that were invented later \cite{Gross, ADF}. 

Since $N = 2n-1$ is odd, the integer $n$ serves 
as the multiplicative inverse of $2$ among the integers modulo $N$. 
It is then convenient to write the Heisenberg group elements as 

\begin{equation} D(i,j) = q^{nij}X^iZ^j \hspace{5mm} \Rightarrow 
\hspace{5mm} D(i,j)D(k,l) = q^{n(jk-il)}D(i+k, j+l) = q^{jk-il}D(k,l)D(i,j) \ . \end{equation}  

\noindent Let us also introduce explicit matrix representations of 
the group generators:

\begin{equation} D(i,j) = q^{nij + bj}\delta_{a,b+i} \ , 
\hspace{10mm} A = \delta_{a+b,0} \ . \end{equation}

\noindent Note that the spectrum of the involution $A$ consists of 
$n$ eigenvalues $1$ and $n-1$ eigenvalues $-1$. Hence $A$ splits the 
vector space into the direct sum 

\begin{equation} {\cal H}_N = {\cal H}_n^{(+)} \oplus {\cal H}_{n-1}^{(-)} 
\ . \end{equation}

\noindent It is these subspaces that we should watch. In fact there 
are altogether $N^2$ subspaces of dimension $n$ singled out in this 
way, because there are $N^2$ involutions 

\begin{equation} A_{ij} = D(i,j)AD(i,j)^\dagger 
\ . \label{Aij} \end{equation} 

\noindent 
The eigenvectors of the various cyclic subgroups can be collected 
into the $N+1$ MUB

\begin{equation} \triangle_{am}^{(k)} = \left\{\begin{array}{cll} 
\delta_{am} & , & k = 0 \\ \ \\ \frac{1}{\sqrt{N}}q^{\frac{(a-m)^2}{2k}} & 
, & 1 \leq k \leq N-1 \\ \ \\ 
\frac{1}{\sqrt{N}}q^{am} & , & k = \infty \end{array} \right. \ 
.\end{equation}

\noindent Here $k$ labels the basis, $m$ the vectors, and $a$ their 
components. For $N = 3$ this coincides with form (\ref{MUB3}) given 
earlier. Note that $N-1$ MUB have been written as circulant matrices, 
which is a convenient thing to do. 
 
The key observation is that the zeroth columns in the MUB all obey---we 
suppress the index labelling components---

\begin{equation} A\triangle_0^{(k)} = \triangle_0^{(k)} \ . \end{equation} 

\noindent Hence this set of $N+1$ vectors belongs to the $n$-dimensional 
subspace ${\cal H}_n^{(+)}$ defined by the involution $A$. We can go on to 
show that each of the $n$-dimensional eigenspaces defined by the 
$N^2$ involutions $A_{ij}$ contain $N+1$ MUB vectors. Conversely, 
each MUB vector belongs to $N$ subspaces. We have found the Segre configuration 

\begin{equation} \left( N(N+1)_N, N^2_{N+1} \right) \end{equation} 

\noindent containing $N^2 + N$ points and $N^2$ $(n-1)$-planes in 
projective $(N-1)$-space, always assuming that $N$ is an odd prime.   

The intersection properties of the Segre configuration are remarkable. 
Two $n$-spaces in $2n-1$ dimensions intersect at least in a single ray. 
With a total of $N^2$ such subspaces to play with we expect many vectors 
to arise in this way. But let $\psi$ be such a vector. A minor calculation 
shows that 

\begin{equation} \psi  = A_{ij}A_{kl}\psi  = 
q^{2(il-jk)}D(2i-2k, 2j-2k)\psi  \ . \end{equation}

\noindent Thus $\psi$ must be an eigenvector of some element in the Heisenberg 
group, and hence the intersection of any two $n$-spaces is always one 
of the $N(N+1)$ eigenvectors in the configuration. In the other direction 
things are a little more complicated. Two vectors belonging to the same 
basis are never members of same eigenspace, while two vectors of two 
different MUB belong to a unique common eigenspace. Using 
projective duality we obtain the dual configuration 

\begin{equation} \left( N^2_{N+1}, N(N+1)_N \right) \end{equation} 

\noindent consisting of $N^2$ $(n-1)$-spaces and $N^2 + N$ hyperplanes. 
The intersection properties are precisely those of a finite affine plane 
\cite{Dolgachev}. 

These are the facts that so delighted Segre. A hundred years 
later they delighted Wootters \cite{Wootters1}---although he phrased the 
discussion directly in terms of the phase point 
operators $A_{ij}$ rather than in terms of their eigenspaces.  
A systematic study of prime power dimensions in Segre's spirit appears 
not to have been made, although there are some results for $N = 9$ \cite{Horadam}. 

But where is the SIC? It is hard to tell. When the dimension 
$N = 2n-1 = 3$ we observe that $n-1 = 1$, so the dual Segre configuration 
involves $N^2$ vectors, and these are precisely the SIC vectors (\ref{points}). 
When $N \geq 5$ the Segre configuration contains not even a candidate 
set of $N^2$ vectors. But at least, as a byproduct of the construction, 
we find a set of $2n$ equiangular vectors in any $n$ dimensional Hilbert 
space such that $2n-1$ is an odd prime. Explicitly they are  

\begin{equation} \left[ \begin{array}{ccccc} \sqrt{2n-1} & 1 & 1 & \dots 
& 1 \\ 0 & \sqrt{2} & \sqrt{2}q^{1\cdot 1^2} & \dots & \sqrt{2}q^{(2n-2)\cdot 1^2} \\  
0 & \sqrt{2} & \sqrt{2}q^{1\cdot2^2} & \dots & \sqrt{2}q^{(2n-2)\cdot 2^2} \\ 
\vdots & \vdots & \vdots & & \vdots \\ 
0 & \sqrt{2} & \sqrt{2}q^{1\cdot (n-1)^2} & \dots & \sqrt{2}q^{(2n-2)(n-1)^2} 
\end{array} \right] \ . \label{2n} \end{equation}    

\noindent Such sets are of some interest in connection with pure state 
quantum tomography \cite{Flammia}. 

The elliptic curve itself has not been much in evidence in this section. 
It is still there in the background though, and in any dimension it 
contains $N^2$ distinguished $N$-torsion points. A study of the explicit 
expression for the Heisenberg covariant elliptic curve shows that each 
of its torsion points belong to one of the $N^2$ eigenspaces 
${\cal H}_{n-1}^{(-)}$ \cite{Hulek}, and with the single exception of the 
$N = 3$ example (\ref{points}) the known SICs never sit in such a subspace, 
so the torsion points are not SICs. This is discouraging, but we will 
find some consolation when we proceed to examine the $N = 4$ case. 
 
\section{The SIC in 4 dimensions}
In an $N = 4$ dimensional Hilbert space there is a parting of the ways, 
in the sense that the MUB and the SIC are defined using two different 
versions of the Heisenberg group. The elliptic curve stays with $H(4)$. 
Using an argument concerning line bundles and employing ingredients such 
as the Riemann-Roch theorem, it can be shown that an elliptic 
normal curve in projective 3-space (not confined to any projective plane) 
is the non-singular intersection of two quadratic polynomials. If we insist 
that it is transformed into itself by the Heisenberg group in its clock 
and shift representation (\ref{group}), it follows \cite{Hulek} that 
these quadratic polynomials are  

\begin{equation} Q_0 = x_0^2 + x_2^2 + 2ax_1x_3 \ , \hspace{8mm} Q_1 = 
x_1^2 + x_3^2 + 2ax_0x_2 \ . \end{equation}

\noindent The extra symmetry under the involution $A$, defined in (\ref{A}), 
again appears 
automatically. We can diagonalise these quadratic forms by means of a unitary 
transformation of our Hilbert space. In the new coordinates we have 

\begin{equation} Q_0 = z_0^2 + iz_1^2 + a(iz_2^2 + z_3^2) \ , \hspace{8mm} 
Q_1 = iz_2^2 - z_3^2 + a(z_0^2 - iz_1^2) \ . \end{equation}

\noindent Note that $Q_0 = Q_1 = 0$ implies 

\begin{equation} z_0^4 +z_1^4 + z_2^4 + z_3^4 = 0 \ . \end{equation}

\noindent Hence the elliptic curve lies on a quartic surface.  

The new basis that we have introduced has a natural interpretation in 
terms of the involution $A$. First of all, by acting on $A$ with the 
Heisenberg group as in eq. (\ref{Aij}) we obtain only four involutions 
altogether, rather than $N^2$ as in the odd prime case. Their spectra 
are $(1,1,1,-1)$, and in the new basis they are all represented by 
diagonal matrices. Hence each basis vector is inverted by one involution, 
and left invariant by the others. In projective 3-space they correspond 
to four reference points, and one can show that the 16 tangents of the 16 torsion 
points on the curve divide into 4 sets of 4 each coming together at one of 
the 4 reference points \cite{Hulek}. Each such set is an orbit under the 
subgroup of elements of order 2. 

In our preferred basis the generators of the Heisenberg group appear in the form 

\begin{equation} Z = e^{\frac{i\pi}{4}} \left( \begin{array}{rrrr} 0 & 1 & 0 & 0 \\ 
-i & 0 & 0 & 0 \\ 
0 & 0 & 0 & -i \\ 0 & 0 & -1 & 0 \end{array}\right) \ , \hspace{9mm} 
X = e^{\frac{i\pi}{4}} \left( \begin{array}{rrrr} 
0 & 0 & 1 & 0 \\ 0 & 0 & 0 & 1 \\ -i & 0 & 0 & 0 \\ 0 & i & 0 & 0 \end{array} 
\right) \ . \end{equation}

\noindent Finding a set of 16 SIC-vectors covariant under the Heisenberg group 
is now a matter of simple guesswork. One answer, ignoring overall phases and 
normalisation, is   

\begin{equation} \left[ 
\begin{array}{rrrrrrrrrrrrrrrr} x & x & x & x & i & i & - i & - i & i &  i 
& - i & - i & i & i & - i & - i \\ 1 & 1 & - 1 & - 1 & x & x & x & x & i & -i 
& i & - i & 1 & - 1 & 1 & - 1 \\ 1 & -1 & 1 & -1 & 1 & - 1 & 1 & -1 & x & 
x & x & x & - i & i & i & - i \\ 
1 & - 1 & - 1 & 1 & -i & i & i & - i & - 1 & 1 & 1 & - 1 & x & x & x & x 
\end{array} \right] \ , \label{SIC4} \end{equation}

\noindent where 

\begin{equation} x = \sqrt{2 + \sqrt{5}} \ . \end{equation}

\noindent All scalar products have the same modulus because 

\begin{equation} (x^2-1)^2 = |x+1 + i(x-1)|^2 \ . \end{equation} 

\noindent Thanks to our change of basis, this is significantly more memorable 
than the standard solutions \cite{Zauner, Renes} (and it was in fact arrived 
at, without considering the Heisenberg group at all, by Belovs \cite{Belovs}). 
The whole set is organised 
into 4 groups, where each group sits at a standard distance from the 4 basis 
vectors that are naturally singled out by the elliptic curve. The normalised vectors 
obey eq. (\ref{MUS}) for a Minimum Uncertainty State, even though our basis 
is unusual. 

The otherwise mysterious invariance of the SIC vectors under some element 
of the Clifford group of order 3 is now easy to see. We focus 
on the group of vectors 

\begin{equation} \left[ \begin{array}{rrrr} x & x & x & x \\ 1 & 1 & -1 & -1 \\ 
1 & -1 & 1 & -1 \\ 1 & -1 & -1 & 1 \end{array} \right] \ . \end{equation}

\noindent They form an orbit under the subgroup of elements of order 2. When 
we project them to the subspace orthogonal to the first basis vector we have 
4 equiangular vectors in a 3 dimensional subspace. Each projected vector 
will be invariant under a rotation of order 3 belonging to the symmetry 
group of this tetrahedron. An example leaving the first vector invariant is 

\begin{equation} R = \left( \begin{array}{cccc} 1 & 0 & 0 & 0 \\ 0 & 0 & 0 & 1 \\ 
0 & 1 & 0 & 0 \\ 0 & 0 & 1 & 0 \end{array} \right) \ . \end{equation}

\noindent It is straightforward to check that the rotation $R$ belongs to 
the Clifford group, and is indeed identical to one of "Zauner's unitaries" 
\cite{Zauner}.

Each of the four involutions $A$ admit a "square root" belonging to the 
Clifford group, such as 

\begin{equation} F = \left( \begin{array}{cccc} 1 & 0 & 0 & 0 \\ 0 & 0 & 1 & 0 \\ 
0 & 1 & 0 & 0 \\ 0 & 0 & 0 & i \end{array} \right) \hspace{5mm} 
\Rightarrow \hspace{5mm} F^2 = A = \left( \begin{array}{cccc} 1 & 0 & 0 & 0 \\ 
0 & 1 & 0 & 0 \\ 
0 & 0 & 1 & 0 \\ 0 & 0 & 0 & -1 \end{array} \right)\ . \end{equation}

\noindent Acting with these unitaries on the SIC (\ref{SIC4}) will give a 
set of altogether 16 different SICs, collectively forming an orbit under the 
Clifford group \cite{Zauner, Marcus}. 

Note that the 16 SIC points in projective space do not actually sit on 
the elliptic curve. In this sense the step from $N = 3$ to $N = 4$ is 
non-trivial. In an arbitrary even dimension $N = 2n$ 
the involution $A$, see (\ref{A}), has a spectrum consisting of $n+1$ 
eigenvalues $1$ and $n-1$ eigenvalues $-1$. When $N = 4$ this singles out 
a unique ray but in higher dimensions this is not so, so generalising 
to arbitrary even dimension will not be easy. 

\section{Minimum Uncertainty States in four dimensions}
Eddington and his surface have not yet appeared. The group on whose twofold 
cover his Fundamental Theory hinged was not the Heisenberg group over 
the ring of integers modulo 4, but a different Heisenberg group of 
the form $H(2)\otimes H(2)$ \cite{Eddington}. This group can be 
represented by real matrices, and is in fact the group which gives rise 
to the complete set of MUB in 4 dimensions. What can we do with it?

There does not exist a SIC which is covariant under Eddington's group. 
In fact the group $H(2)^{\otimes k}$ admits such an orbit only if 
$k = 1$ or $k = 3$ \cite{Godsil}.  
As a substitute we can look for an orbit of 16 Minimum Uncertainty 
States with 
respect to the maximal set of MUB. Such an orbit does exist, and is 
given by the 16 vectors 

\begin{equation} \left[ \begin{array}{cccccccccccccccc} x & x & x & x & 
\alpha & \alpha & - \alpha & - \alpha & \alpha & \alpha & - \alpha & 
- \alpha & \alpha & \alpha & - \alpha & - \alpha \\ 
\alpha & \alpha & - \alpha & - \alpha & x & x & x & x & 
\alpha & - \alpha & \alpha & - \alpha & \alpha & 
- \alpha & \alpha & - \alpha \\ 
\alpha & - \alpha & \alpha & - \alpha & \alpha & - \alpha & \alpha 
& - \alpha & x & x & x & x & \alpha & -\alpha & - \alpha & \alpha \\ 
\alpha &  -\alpha & - \alpha & \alpha & \alpha & - \alpha & - \alpha 
& \alpha & \alpha & - \alpha & - \alpha & \alpha & x & x & x & x 
\end{array} \right] \ , \label{Edd} \end{equation}

\noindent where 

\begin{equation} x = \sqrt{2 + \sqrt{5}} \ , \hspace{8mm} \alpha = e^{ia} \ ,  
\hspace{8mm} \cos{a} = \frac{\sqrt{5}-1}{2\sqrt{2 + \sqrt{5}}} \ . \end{equation}

\noindent I omit the lengthy proof that these 16 vectors really are 
Minimum Uncertainty States \cite{Asa}. Although this is not a SIC, 
in a way it comes close to being one. Like the SICs (\ref{points}) and 
(\ref{SIC4}), it can be arrived at using the following procedure: Introduce a 
vector $(x, e^{i\mu_1}, \dots , e^{i\mu_{N-1}})^{\rm T}$, and adjust the value 
of $x$ so that the normalised vector solves eq. (\ref{MUS}) for a Minimum 
Uncertainty State. Next introduce a complex Hadamard matrix, that is to 
say a unitary matrix all of whose matrix elements have the same modulus. 
Such matrices exist in any dimension, although their classification problem 
is unsolved if the dimension exceeds 5 \cite{Tadej}. By multiplying with 
an overall factor $\sqrt{N}$, and then multiplying the columns with phase 
factors, we can ensure that all matrix elements in the first row equal 
1. Replace these elements with $x$. Next multiply the rows 
with phase factors until one of the columns equals the vector we introduced. 
The result is a set of $N$ vectors with all mutual scalar products taking 
the value that characterises a SIC. Next permute the entries of the original 
vector cyclically, and afterwards try to adjust the phases $\mu_a$ so that the 
resulting $N$ vectors are again equiangular with the mutual scalar products 
characterising a SIC. Extending the new vectors using an Hadamard matrix 
in the same way as before then gives $N$ equiangular vectors each of which 
belongs to a separate group of $N$ equiangular vectors. Before we can say 
that we have constructed a SIC we must check that all scalar products 
between pairs of vectors not belonging to the same group take the SIC values. 
The vectors (\ref{Edd}) fail to form a SIC only because the last step fails.  

Finally we come back to Eddington's lecture room. In the treatise that 
he read \cite{Hudson} 
it is explained that an orbit of $H(2)\otimes H(2)$ gives a realisation of 
the Kummer configuration $16_6$, consisting of 16 points and 16 planes in 
projective 3-space, such that each point belongs to 6 planes and each plane 
contains 6 points. The above set of Minimum Uncertainty States realises this 
configuration. As an example, the 6 vectors  

\begin{equation} \left[ 
\begin{array}{cccccc} - \alpha & - \alpha & - \alpha & - \alpha & - \alpha 
& - \alpha \\ 
x & x & \alpha & - \alpha 
& \alpha & - \alpha \\ \alpha & -\alpha & x & x & - \alpha & \alpha \\ 
-\alpha & \alpha & - \alpha & \alpha & x & x  
\end{array} \right] \end{equation}

\noindent are orthogonal to the row vector 

\begin{equation} ( \begin{array}{ccccc} x & \alpha & \alpha & 
\alpha \end{array} ) \ , \end{equation}

\noindent or in other words the corresponding 6 points belong to the 
corresponding plane. This is a purely group theoretical property and does 
not require the vectors to be Minimum Uncertainty States. Still, Eddington's 
story suggests that our 16 special vectors may have some use, somewhere.

\section*{Acknowledgments}
I thank Subhash Chaturvedi for telling me about the Segre configuration, at a 
point in time when neither of us knew about Segre. Both of us give our best 
wishes to Tony!

\smallskip

\end{document}